\documentclass[showpacs,twocolumn,floatfix,prl]{revtex4}
\usepackage{graphicx}
\usepackage{amsfonts}
\usepackage{amsmath}
\usepackage{amssymb}
\usepackage{graphicx}%

\newcommand{\beq}{\begin{equation}}
\newcommand{\eeq}{\end{equation}}
\newcommand{\beqn}{\begin{eqnarray}}
\newcommand{\eeqn}{\end{eqnarray}}
\newcommand{\bearr}{\begin{array}}
\newcommand{\enarr}{\end{array}}

\begin{document}

\title{Vortex and translational currents due to broken time-space symmetries}

\author{S. Denisov$^1$, Y. Zolotaryuk$^{2}$, S. Flach$^{3}$,
and O. Yevtushenko$^4$}

\affiliation{$^1$ Institut f\"ur Physik, Universit\"at  Augsburg,
       Universit\"atsstr.1, D-86135 Augsburg, Germany}
\affiliation{$^2$ Bogolyubov Institute for Theoretical Physics,
National Academy of Sciences of Ukraine,
 03680 Kiev, Ukraine}
\affiliation{$^3$Max-Planck-Institute for the Physics of Complex Systems,
N\"othnitzer Str. 38, D-01187 Dresden, Germany}
\affiliation{$^4$ Physics Department, Arnold Sommerfeld Center for Theoretical Physics,
                 Ludwig-Maximilians-Universit{\"a}t M{\"u}nchen, D-80333 M{\"u}nchen, Germany}
\date{\today}
\begin{abstract}
We consider the classical dynamics of a particle in a
$d=2,3$-dimensional space-periodic potential under the influence of
time-periodic external fields with zero mean. We perform a general
time-space symmetry analysis and identify conditions, when the
particle will generate a nonzero averaged translational and vortex
currents. We perform computational studies of the equations of
motion and of corresponding Fokker-Planck equations, which confirm
the symmetry predictions. We address the experimentally important
issue of current control. Cold atoms in optical potentials and
magnetic traps are among possible candidates to observe these
findings experimentally.

\end{abstract}
\maketitle

%
%

The idea of directed motion under the action of an external
fluctuating field of zero mean goes back to Smoluchowski and Feynman
\cite{Smol}. It has been intensively studied in the past decades
again \cite{magna}. It is believed to be connected with the
functioning of molecular motors, and can be applied to transport
phenomena which range from mechanical engines to an electron gas
(see \cite{revs,Reim} and references therein).

The separation of the fluctuating fields into an uncorrelated white
noise term and a time-periodic field was used to perform a symmetry
analysis of the most simple case -  a point-like particle moving in
a one-dimensional periodic potential \cite{Flach1}. It allowed to
systematically choose space and time dependencies of potentials and
ac fields such, that a nonzero dc current is generated. Various
studies of the dynamical mechanisms of rectification have been
reported (e.g. \cite{Den}). Among many experimental reports, we
mention the successful testing of the above symmetry analysis using
cold atoms in one-dimensional optical potentials \cite{ren}. By use
of more laser beams, experimentalists can already fabricate two- and
three-dimensional optical potentials, with different symmetries and
shapes \cite{2D3D}, with the aim of even more controlled stirring of
cold atoms in these setups.

A particle which is moving in a $d=2,3$-dimensional periodic
potential may contribute to a directed current along a certain
direction. At the same time, the particle can perform vortex motion
(which is not possible in a one-dimensional setting) generating a
nonzero average of the angular momentum. Directed translational
currents are supported by unbounded trajectories while vortex
currents may be localized in a finite volume. The question is then,
how can we control a type of the directed motion? To answer this
question, one has to find conditions for an appearance of either
purely vertex or purely translational currents. This is the main
goal of the present work.

In this letter we perform a symmetry analysis of particle motion
in a $d=2,3$-dimensional periodic potential, under the influence of external ac fields.
We identify the symmetries which ensure that either directed currents, or vortex currents,
are strictly zero. Breaking these symmetries one by one allows to control
the particle motion, to generate either directed, or vortex, currents, or both simultaneously.

We consider the dynamics of a classical particle (e.g. an atom of a cold dilute
atom gas, loaded onto a proper optical lattice)  exposed to an
external potential field:
\begin{equation}
\mu \ddot{{\bf r}} + \gamma \dot{\bf r} = {\bf g}({\bf r},t)+
\boldsymbol{\xi}(t) ~,
~~{\bf g}({\bf r},t)=-\boldsymbol{\nabla} U({\bf r},t) ~.\\
\label{1sys}
\end{equation}
Here ${\bf r}=\{x,y,z\}$ is the coordinate vector of the particle,
the parameter $\gamma \ge 0$ characterizes the dissipation strength,
and $\mu \ge 0$ defines the strength of the inertial term
\cite{dissip2}. The force ${\bf g}({\bf r},t)=\{g_\alpha({\bf
r},t)\}$, $\alpha=x,y,z$, is time- and space-periodic:
\begin{equation}
{\bf g}({\bf r},t)={\bf g}({\bf r},t+T)={\bf g}({\bf
r}+\mathbf{L},t). \label{sp_per}
\end{equation}
The absence of a dc bias implies
\begin{equation}
\langle {\bf g} ({\bf r} ,t) \rangle_{\mathbf{L},T}\equiv
\int_{0}^T\int_{\mathbf{L}} {\bf g} ({\bf r} ,t) \, dt \, dx dy dz=0
\label{aver}
\end{equation}
where the spatial integration extends over one unit cell.

The fluctuating force is modeled by a $\delta$-correlated Gaussian white
noise, $\boldsymbol{\xi}(t)=\{\xi_{x},\xi_{y}, \xi_{x}\}$,
$\langle \xi_{\alpha}(t) \xi_{\beta}(t') \rangle = 2\gamma D \delta
(t-t') \delta_{\alpha \beta}$ ($\alpha,\beta=x,y,z$).
Here $D$ is the noise strength. The statistical description of the
system (\ref{1sys}) is provided by the Fokker-Planck
equations (FPE) \cite{risken}:
\begin{eqnarray}
\nonumber
\dot{P}(\bf{r},\mathbf{v},t)&=&\{-\nabla_{\mathbf{r}}\cdot \mathbf{v}+
\nabla_{\mathbf{v}}
\cdot [\gamma
\mathbf{v}-\mathbf{g}(\mathbf{r},t)]+\\
&+&\gamma D \triangle_{\mathbf{v}}\}P(\bf{r},\mathbf{v},t),~\mu=1,
~\mathbf{v} = \dot {\bf r},
\label{fpe_under}\\
\gamma \dot{P}(\bf{r},t)&=-&[\nabla_{\mathbf{r}}\cdot
\mathbf{g}(\mathbf{r},t)+  D
\triangle_{\mathbf{r}}]P(\bf{r},t),~\mu=0 \label{fpe_over}~.
\end{eqnarray}
Each of the linear equations
(\ref{fpe_under}-\ref{fpe_over}) has a unique attractor solution,
$\hat{P}$
which is space and time periodic \cite{risken}.

\emph{Directed  transport.} Let us consider the dc component of the directed current
${\bf j}(t) = \mathbf{v} = \dot {\bf r}$
in terms of the attractor $\hat{P}$:
\begin{eqnarray}
&&{\mathbf{J}}=\langle \mathbf{\mathbf{v}}\cdot
\hat{P}(\mathbf{r},\mathbf{v},t)\rangle_{T,\mathbf{L}},~~\mu=1~,
\label{cur_under}\\
&&{\mathbf{J}}=\frac{1}{\gamma}\langle {\bf g}({\bf r},t) \cdot
\hat{P}(\mathbf{r},t)\rangle_{T,
\mathbf{L}},~~\mu=0~.\label{cur_over}
\end{eqnarray}
The strategy is now to identify symmetry operations which invert the
sign of $\mathbf{j}$, and, at the same time, leave Eq. (\ref{1sys})
invariant. If such symmetries exist, the dc current $\mathbf{J}$
will strictly vanish. Sign changes of the current can be obtained by
either inverting the spatial coordinates, or time (simultaneously
allowing for shifts in the other variables). Below we list all
operations together with the requirements the force ${\bf g}$ and
the control parameters have to fulfill:
\begin{eqnarray}
{\widehat S}_1: {\bf r} \to - {\bf r} + {\bf r'},
~t \to t+{\bf \tau};\; ~~~ {\widehat S}_1 ({\bf g}) \to -{\bf g}\;,~~~~~~~\label{S1}\\
{\widehat S}_2: {\bf r} \to {\bf r} + {\bf \Lambda},~t \to
-t+t';~~~~~~~~~~~~~~~~~~~~~~~~~~~~~\nonumber
\\
~~~~~{\widehat S}_2({\bf g}) \to {\bf g}~ (\mbox{if }\gamma=0); ~~~
{\widehat S}_2({\bf g}) \to -{\bf g}~ (\mbox{if }\mu=0).~~~~
\label{S2}
\end{eqnarray}
Here $t'$ and ${\bf r'}$ depend on the particular shape of ${\bf
g}({\bf r},t)$. The system must be invariant under a spatial
translation by the vector $2 {\bf \Lambda}$ in space and $2{\bf
\tau}$ in time, respectively. The vector ${\bf \Lambda}$ is
therefore given by ${\bf \Lambda}=\sum_{\alpha} n_{\alpha} {\bf
L}_\alpha/2$, $n_\alpha=0,1$, while $\tau = 0,T/2$. By a proper
choice of ${\bf g}$ all relevant symmetries can be broken, and one
can then expect the appearance of a non-zero dc current $\mathbf{J}$
\cite{Supersym}.

%
\begin{figure}[t]
\includegraphics[width=7cm,height=4.4cm,angle=0]{f1a.eps}
\includegraphics[width=7.5cm,height=4.4cm,angle=0]{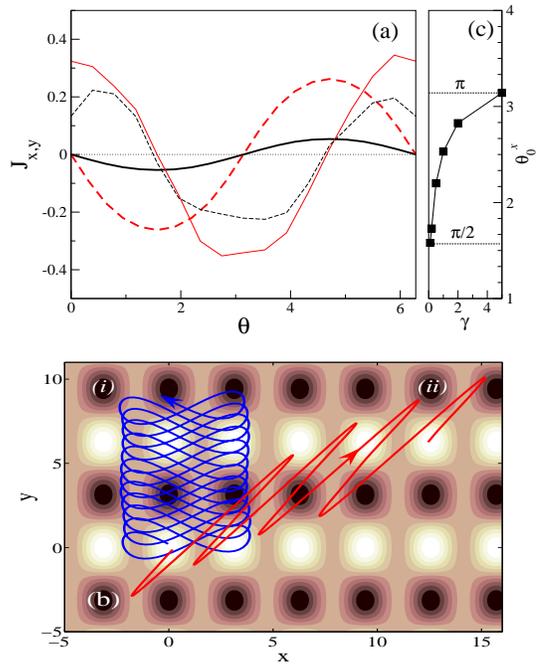}
\caption{(color online) (a) Dependence of the current components, $
J_{x}$(solid line) and $J_{y}$(dashed line), on $\theta$ for
(\ref{1sys}),(\ref{potent1})-(\ref{force1}), with $D=1$,
$E_{x}^{(1)}=-E_{x}^{(2)}=2$, $E_{y}^{(1)}=-E_{y}^{(2)}=2.5$. Data
are for the overdamped ($ \mu = 0, \gamma = 1 $, thick lines) and
underdamped ($ \mu = 1, \gamma = 0.1 $, thin lines) cases,
respectively; (b) The time evolution of the mean particle position,
$\bar{\mathbf{r}}(t) = \int {\mathbf r} P({\mathbf r}, {\mathbf v},
t) d{\mathbf r} d{\mathbf v}$, for $\theta=0$. The trajectories are
superimposed on the contour plot of the potential (\ref{potent1}).
Curve (\textit{i}) corresponds to $E_x^{(1)}=3$,
$E_x^{(2)}=E_y^{(1)}=0$, $E_y^{(2)}=3.5$ and curve (\textit{ii}) -
to the parameters of panel (a). The other parameters are: $ D = \mu
= 1, \gamma = 0.1 $; (c) The phase lag $\theta^{(0)}_x$ as a
function of the dissipation strength $\gamma$. } \label{fig:conv1}
\end{figure}

To be more precise, we consider the case of a particle moving in a
two-dimensional periodic potential and being driven by an external
ac field: ${\bf g}({\bf r},t)=-\boldsymbol{\nabla}V({\bf r})
+\mathbf{E}(t)\equiv \mathbf{f}({\bf r}) +\mathbf{E}(t)$. The
symmetry ${\widehat S}_1$ holds if the potential force is
\textit{anti-symmetric},
$\mathbf{f}(-\mathbf{r}+\mathbf{r}')=-\mathbf{f}(\mathbf{r})$,
 and the driving function is \textit{shift-symmetric},
$\mathbf{E}(t+T/2)=-\mathbf{E}(t)$. The symmetry ${\widehat S}_2$
holds at the Hamiltonian limit, $\gamma=0$, if the driving force is
\textit{symmetric}, $\mathbf{E}(t+t')=\mathbf{E}(-t)$. Finally, the
symmetry ${\widehat S}_2$ holds at the overdamped limit, $\mu=0$ ,
if the potential force is shift-symmetric,
$\mathbf{f}(\mathbf{r}+\mathbf{\Lambda})=-\mathbf{f}(\mathbf{r})$
and the driving force is anti-symmetric,
$\mathbf{E}(t+t')=-\mathbf{E}(-t)$.

In order to break the above symmetries, we choose
\begin{eqnarray}
&&V({\bf r})=V(x,y)=\cos(x)[1+\cos(2 y)]~, \label{potent1}\\
&&E_{x,y}(t)=E_{x,y}^{(1)}\sin  t  +E_{x,y}^{(2)}\sin (2 t +\theta)~.
        \label{force1}
\end{eqnarray}
The potential (\ref{potent1}) is shift-symmetric,
$\mathbf{\Lambda}=\{\pm\pi,0\}$. The symmetry ${\widehat S}_1$
is broken since $E$ is not shift-symmetric. Therefore in general we
expect $\textbf{J} \neq 0$.

In Fig.\ref{fig:conv1} we show the computational evaluation of
equations (\ref{fpe_under},\ref{fpe_over}) \cite{numeric}. We
confirm the presence of a nonzero dc current. Applying operations
${\widehat S}_1$ and $\theta \to \theta+\pi$ we conclude
$\textbf{J}(\theta+\pi)=-\textbf{J}(\theta)$, which allows for an
easy inversion of the current direction, as also confirmed by the
data in Fig.\ref{fig:conv1}(a). In the overdamped limit $\mu=0$,
${\widehat S}_2$ is restored for $\theta=0,\pm\pi$, and therefore
$\textbf{J}(-\theta)=-\textbf{J}(\theta)$ (thick lines in
Fig.\ref{fig:conv1}(a)). Upon approaching the Hamiltonian limit,
$\gamma \rightarrow 0$, the points where ${\bf J}=0$ shift from
$\theta=0,\pi$ to $\theta=\pm \pi/2$ where the symmetry ${\widehat
S}_2$ is restored (thin lines in Fig.\ref{fig:conv1}(a)). In the
underdamped regime, the dc-current can be approximated as $J_\alpha
\propto  J_\alpha^{(0)} \sin[\theta-\theta_\alpha^{(0)}(\gamma)],
~\alpha=x,y$. The phase lag is equal to $\theta^{(0)}_{x,y}= \pi/2$
and  $\theta^{(0)}_{x,y}= 0$ in the Hamiltonian and overdamped
limits, respectively (Fig.\ref{fig:conv1}(c)) \cite{Renzoni_dissip}.

Even more control over the current direction is possible, by
imposing the symmetry conditions (\ref{S1}-\ref{S2}) on each
component $g_{\alpha}(\mathbf{r},t)$ independently. For
(\ref{potent1}) and (\ref{force1}) with $E_x^{(2)}=E_y^{(1)}=0$,
$\theta=0$ the symmetry transformation $ {\widehat S}_c~:~~ x \to -
x,~~~ y \to y,~~~t \to t+\pi~$ implies that the current along the
$x$-direction is absent, $J_x=0$, and directed transport is
happening along the $y$-axis [see Fig.1b, curve (i)]. We may
conclude, that the symmetry analysis turns out to be a powerful tool
of predicting and controlling directed currents of particles which
move in $d=2,3$-dimensional potentials under the influence of
external ac fields. Note that dynamical mechanisms of current
rectification of a two-dimensional deterministic tilting ratchet
were discussed in Ref. \cite{guano_apes}.

%
%
{\it Vorticity.}  At variance to the one-dimensional case,
particles in two and three dimensions can perform vortex motion, thereby
generating ring currents, or nonzero angular momentum.
First of all we note that the particle dynamics is not confined to one
spatial unit cell of the periodic potential $U(\mathbf{r},t)$. Even in the
case when a directed current is zero due to the above symmetries, $\mathbf{J}=0$, the particle
can perform unbiased diffusion in coordinate space. In order to distinguish between
directed transport
and spatial diffusion on one side, and rotational currents on the other side, we
use the angular velocity
\cite{k91}
\begin{equation}
{\bf \Omega}(t)= [\dot {\bf r}(t) \times \ddot {\bf r}(t)]/\dot {\bf r}^2(t),
 ~~{\bf
J}_{\Omega}=\langle {\bf \Omega}(t) \rangle_t \label{angular2}~,
\end{equation}
as a measure for the particle rotation, where
$\langle...\rangle_{t}=\lim_{t\rightarrow
\infty}\frac{1}{t}\int_{0}^{t}... dt'$. ${\bf \Omega}(t)$ is
invariant under translations in space and time. It describes the
speed of rotation with which the velocity vector ${\bf {\dot r}}$
(the tangential vector to the trajectory ${\bf r}(t)$) encompasses
the origin.

Using the above strategy, we search for symmetry operations which
leave the equations of motion invariant, but do change the sign of
the angular velocity. If such symmetries exist, rotational currents
strictly vanish on average. The sign of $\mathbf{\Omega}$ can be
inverted by either \textit{(i)} time inversion $t\rightarrow -t$
together with an optional space inversion $\mathbf{r} \rightarrow
\pm\mathbf{r}$, or \textit{(ii)} the permutation of any two
variables, e.g. ${\widehat {\cal P}_{xy}}: \{x,y,z \} \to \{y,x,z
\}$. That leads to the following possible symmetry transformations:
\begin{eqnarray}
{\widehat R}_1~: {\bf r} \to  {\widehat {\cal P}}{\bf r} +
 {\bf r'},~~t \to t+\tau\;;\;~~
{\widehat R}_1(\mathbf{g}) \to \mathbf{g}\;,\;~~~~\label{R1}\\
{\widehat R}_2: {\bf r} \to {\bf \pm r} + {\bf \Lambda},~~t \to
-t+t',\; ~~~~~~~~~~~~~~~~~~~~~~~\nonumber \\
~~{\widehat R}_2({\bf g}) \to {\bf g} ~(\mbox{if }\gamma=0);
~~{\widehat R}_2({\bf g}) \to -{\bf g} ~(\mbox{if
}\mu=0).~~~\label{R2}
\end{eqnarray}
Here ${\widehat {\cal P}}$ stands for any of the following
operations: ${\widehat {\cal P}_{xy}}$, ${\widehat {\cal P}_{yz}}$
or ${\widehat {\cal P}_{zx}}$ and $t'$ and ${\bf r'}$ again depend
on the particular shape of $ {\bf g}( {\bf r}, t) $.

To be concrete, we will again consider a particle moving in a
two-dimensional periodic potential and being driven by an external
ac field: ${\bf g}({\bf r},t)=-\boldsymbol{\nabla}V({\bf r})
+\mathbf{E}(t)\equiv \mathbf{f}({\bf r}) +\mathbf{E}(t)$. For $d=2$
there is an additional transformation due to a mirror reflection at
any axis, $\hat{\Sigma}_x:\{x,y\}\rightarrow \{x,-y\}$ or
$\hat{\Sigma}_y:\{x,y\}\rightarrow \{-x,y\}$,
\begin{equation}
{\widehat R}_3~:~~ {\bf r} \to  \hat{\Sigma} {\bf r}~,~t \to
t+T/2;~~~ {\widehat R}_3(\mathbf{g}) \to \mathbf{g}\ ~~ \label{R4}
\end{equation}
Symmetry ${\widehat R}_1$ can be satisfied for the {\it
Hamiltonian}, {\it underdamped} and {\it overdamped} cases if
${\widehat {\cal P}}{\bf f}({\widehat {\cal P}}{\bf r}+{\bf r'})
={\bf f}({\bf r})$ and ${\widehat {\cal P}}{\bf E}(t+t')={\bf
E}(t)$. The symmetry ${\widehat R}_{2}$ apply for the same cases as
their counterparts ${\widehat S}_{2}$ if there is no space
inversion. In the presence of space inversion they can be satisfied
both in the Hamiltonian [if ${\bf f}(-{\bf r})=-{\bf f}({\bf r})$
and ${\bf E}(t+t')=-{\bf E}(-t)$] and {overdamped} [if ${\bf
f}(-{\bf r})={\bf f}({\bf r})$ and ${\bf E}(t+t')={\bf E}(-t)$]
limits. The symmetry ${\widehat R}_3$ is relevant if $ \hat\Sigma_x
$ can be applied: $ f_x(x,-y)=f_x(x,-y)$, $f_y(x,-y)=-f_y(x,-y)$,
$E_x(t+T/2) =E_x(t)$ and $E_y(t+T/2)=-E_y(t)$. Similar conditions
can be found for $ \hat\Sigma_y $.

%
\begin{figure}[t]
\includegraphics[width=8cm,height=4.7cm,angle=0]{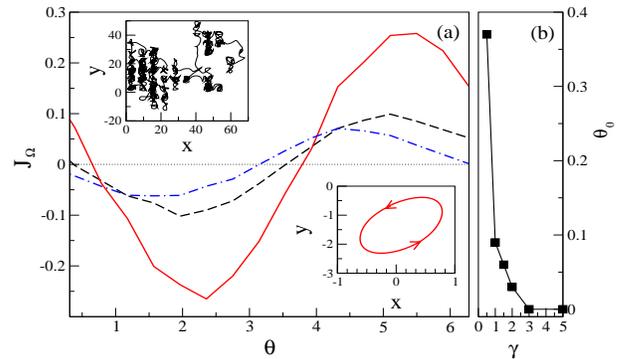}
\caption{(color online) (a) Dependence $J_{\Omega}(\theta)$,
Eq.(\ref{angular2}), for (\ref{1sys}),
(\ref{potent3})-(\ref{force3}), with $\mu=1$, $D=0.5$,
$E_x^{(1)}=0.4$, $E_y^{(1)}=0.8$ and $\gamma=0.2$ (solid line),
$\gamma=0.05$ (dashed line), and $\gamma=2$ (dashed-dotted line).
Insets: the trajectory (left insert) and the corresponding attractor
solution, $\bar{\mathbf{r}}(t)$, (right inset) for the case
$\gamma=0.2$ and $\theta=\pi/2$. We have used $N=10^{5}$ independent
stochastic realizations to perform the noise averaging; (b) The
phase lag $\theta_0$ as a function of the dissipation strength
$\gamma$.} \label{fig:conv3}
\end{figure}

We performed numerical integrations of the equation of motion (\ref{1sys})
with the following potential
and driving force:
\begin{eqnarray}
&&V(x,y)=[-3\left (\cos x+\cos y\right )+  \cos x \cos y]/2,~~ \label{potent3}\\
&&E_x(t)=E_x^{(1)}\cos t~,~~ E_y(t)=E_y^{(1)}\cos(t+\theta)~.
\label{force3}
\end{eqnarray}
Averaging was performed over $N=10^5$ different stochastic
realizations \cite{problems}. Fig.\ref{fig:conv3} shows the
dependence of the rotational current (\ref{angular2}) on the
relative phase $\theta$. The system is invariant under the
transformation ${\widehat S}_1$ (\ref{S1}), therefore the directed
current ${\bf J}=0$. However, for the underdamped case, $\gamma \neq
0$, all the relevant symmetries (\ref{R1}-\ref{R4}) are violated,
and the resulting rotational current (\ref{angular2}) is nonzero,
and depends on the phase $\theta$ (Fig.2a).  Note that symmetry
${\widehat R}_2$ is restored when $\theta=0,\pm \pi$, thus the
current disappears in the Hamiltonian and overdamped limits for
these values of the phase. The left upper inset in
Fig.\ref{fig:conv3} shows the actual trajectory of a given
realization, confirming that the particle is acquiring an average
nonzero angular momentum, while not leaving a small finite volume
due to slow diffusion and absence of directed currents.

The exact overdamped limit, $\mu=0$, is singular for the definition
(\ref{angular2}) since the velocity of a particle, $ {\bf \dot
r\textrm{(t)}}$, is a nowhere differentiable function. The
overdamped limit can be approached by increasing $\gamma$ at a fixed
$ \mu = 1 $. Alternatively, one may remove the restriction on $ \mu
$ allowing for an infinitesimal value $ 0 < \mu \ll 1 $ at a fixed
dissipation strength $ \gamma = 1 $. Both parameter choices equally
regularize (\ref{angular2}). Numerical simulations for the former
way of regularization show that if $ \gamma/\mu \geq 5$ then the
rotational current completely reflects the symmetries corresponding
to the overdamped case (see dependence $\theta_0(\gamma)$ in
Fig.\ref{fig:conv3}(b)).

Let us discuss the relation of our
results to the case of multidimensional stochastic tilting ratchets
under the influence of a colored noise studied previously \cite{ghosh}. Since
equivalent (in a statistical
sense) stochastic processes, $\boldsymbol{\xi}(t)$, have
been used as driving forces, the symmetry ${\widehat R}_1$
(\ref{R1}) can be violated only by an asymmetric potential.
But all potentials considered in Refs. \cite{ghosh} are invariant
under the permutation transformation ${\widehat {\cal P}}$. As a
consequence, vortex structures for a \textit{local velocity field}
presented in Refs. \cite{ghosh} are completely symmetric
(clockwise vortices are mapped into counterclockwise ones by
${\widehat {\cal P}}$) and, therefore, the average rotation
for any trajectory equals zero.

The phase space dimension is five for $d=2$ and seven for $d=3$.
Therefore, in the Hamiltonian limit ($\gamma=0$), Arnold diffusion
\cite{arnold} takes place. The particle dynamics is no longer
confined within chaotic layers of finite width. That leads to
unbounded, possibly extremely slow, diffusion in the momentum
subspace via a stochastic web \cite{arnold}. Therefore a direct
numerical integration of the equations of motion may lead to
incorrect conclusions.

There is rich variety of physical systems, where multidimensional
ratchets can be observed: cold atoms in two- and three-dimensional
potentials (optical guiding) \cite{hagman}, colloidal particles on
magnetic bubble lattices \cite{tierno}, ferrofluids
\cite{emrj03prl}, and vortices in superconducting films with pinning
sites \cite{souza}. Our methods of directed current control can be
used for an enhancement of the particle separation in the laser
beams of complex geometry \cite{Dholakia}.

To conclude, we formulated space-time symmetries for the absence of
both directed currents, and rotational currents, for particles
moving in spatially periodic potentials, under the influence of
external ac fields. Proper choices of these potentials and fields
allow to break the above symmetries, and therefore to generate and
control directed and rotational vortex currents in an independent
way. Numerical studies supplement the symmetry analysis and confirm
the conclusions.

{\it Acknowledgments.} This work has been partially
 supported by the
DFG-grant HA1517/31-1 (S. F. and S. D.), National Academy of
Sciences of Ukraine through the special program for young scientists
(Y.Z.). O.Y. acknowledges support from SFB-TR-12.


\begin{thebibliography}{10}
\bibitem{dissip2} The underdamped regime is given by $\mu=1$, $\gamma > 0$,
and the Hamiltonian and overdamped limits by $\gamma=0$ and $\mu=0$, correspondingly.
\bibitem{numeric} We have solved the FPE in the underdamped case,
Eq.(\ref{fpe_under}), using a split-step method  \cite{spectral}.
FPE in the overdamped limit, Eq.(\ref{fpe_over}), has been solved
using the Fourier expansion over spatial variables \cite{Den}.
\bibitem{problems}
The Fokker-Planck equations are not useful here, since the angular
velocity involves the acceleration. Therefore one needs to evaluate
averages of products of dynamical variables and white noise terms.
\end{thebibliography}
\end{document}